\documentclass[pra,twocolumn,superscriptaddress]{revtex4-1}
\usepackage{float,graphicx,multirow, amsmath,color,dsfont}
\usepackage{amsmath,bm}
\usepackage{amssymb,mathrsfs,dsfont}
\usepackage{amssymb,mathbbol}
\usepackage{amsfonts}
\usepackage{mathrsfs}
\usepackage{color,hyperref}
\definecolor{darkblue}{rgb}{0.0,0.0,0.3}
\hypersetup{colorlinks,breaklinks,
	linkcolor=darkblue,urlcolor=darkblue,
	anchorcolor=darkblue,citecolor=
	darkblue,pdfauthor=Chandan, pdftitle=ChandanStirling }

\newcommand{\ie}{\textit{i}.\textit{e}.}

\usepackage[normalem]{ulem}
\usepackage{ulem}

\begin{document}
	\title{Temperature dependent maximization of work and efficiency in a degeneracy assisted quantum Stirling heat engine}
	\author{Sarbani Chatterjee}
	\email{mp18015@iisermohali.ac.in}
	\affiliation{Department of Physical Sciences, Indian
		Institute of Science Education and Research (IISER)
		Mohali, Sector 81 SAS Nagar, Manauli P.O. 140306 Punjab, India.}
		\author{Arghadip Koner}
	\email{akoner@ucsd.edu}
	\affiliation{Department of Chemistry and Biochemistry, University of California, San Diego,  9500 Gilman Dr, La Jolla, CA 92093, United States.}
		\author{Sohini Chatterjee}
	\email{schatterjee@jncasr.ac.in}
	\affiliation{Chemistry and Physics of Materials Unit, Jawaharlal Nehru Centre for Advanced Scientific Research (JNCASR), Jakkur P.O. Bangalore, 560064, India. }
		\author{Chandan Kumar}
	\email{ph12129@iisermohali.ac.in}
	\affiliation{Department of Physical Sciences, Indian
		Institute of Science Education and Research (IISER)
		Mohali, Sector 81 SAS Nagar, Manauli P.O. 140306 Punjab, India.}

\begin{abstract}
We propose a quantum Stirling heat engine with an ensemble of harmonic oscillators  as the working medium. We show that the efficiency of the harmonic oscillator quantum Stirling heat engine (HO-QSHE)  at a given frequency can be maximized at a  specific ratio of the temperatures of the thermal reservoirs.  In the low temperature  or equivalently high frequency  limit of the harmonic oscillators, the efficiency of the  HO-QSHE  approaches the Carnot efficiency. Further, we analyse quantum Stirling heat engine  with an ensemble of particle in box  quantum systems as the working medium.  Here both work and efficiency can be maximized at a specific ratio of temperatures of the thermal reservoirs.  These studies will enable us to operate the quantum  Stirling heat engines at its  optimal performance.  The  theoretical study of the HO-QSHE  would provide impetus for its experimental realisation, as most real systems can be approximated as harmonic oscillators  for small displacements near equilibrium.
\end{abstract}
\maketitle
\section{Introduction}
\label{sec:intro}
Thermodynamics started as an exact science at the length scales of macroscopic objects.  The  laws of classical thermodynamics were derived empirically and thus were more robust than any other theory of   that time.   
One of the many practical aspects of this theory was to find the fundamental upper bound in the efficiency of heat engines~\cite{carnot2012reflections,maxwell1860illustration, boltzmann1872wien}, which are devices that utilise the spontaneous heat flow from a hot to a cold bath, and in the process, convert  this heat into mechanical work. Today, the experimental advances in quantum physics have pushed  the use of macroscopic thermodynamics and its partner, classical statistical mechanics, to even smaller length scales~\cite{jan-prx-2017,Peterson-prl-2019}.  These advancements  have led to a successful generalisation of  the classical thermodynamic processes  to their corresponding  quantum versions~\cite{Vinjanampathy,binder2019thermodynamics,amit-arxiv-2020}.
Quantum heat engines are `microscopic' versions of the macroscopic thermodynamic cycles that capitalise on the `quantumness' of the working system to generate positive work~\cite{binder2019thermodynamics}. 

The first proposed quantum heat engine was a three-level maser  which operated with Carnot efficiency in the limiting case~\cite{scovil1959three}. 
Using the generalisation from classical to quantum  theory,  the  theoretical construction of quantum mechanical versions of various classical  engines such as Otto, Carnot, Stirling, Brayton, and Diesel  have been achieved~\cite{PhysRevE.76.031105,mahler-heatengine,PhysRevE.79.041129}. 
A lot of work is in progress  which looks forward to designing and  deriving the optimal performances of quantum mechanical heat engines from microscopic mechanical
laws~\cite{chen2020exergy,tajima2017finite,wang2009performance,stefanatos2014optimal,george-pra-2019,wu2006generalized,PhysRevB.88.214421}. 
 Quantum heat engines  using non-markovian~\cite{zhang2014quantum}, quantum coherent~\cite{scully2003extracting,quan2006quantum}, quantum squeezed~\cite{Huang-pre-2012,lutz-prl-2014,jan-prx-2017,wolfgang-nature-2018,wang-pra-2019},  and entangled~\cite{dillenschneider2009energetics} baths with efficiencies beyond the  classical Carnot limit but with no violations to  the second law of thermodynamics have also been proposed. 
 While  the extracted work and efficiency depend on the working media in a quantum heat engine, these properties are indifferent to the working media used in a classical heat engine.
Different quantum mechanical working media, for instance,  multi-level quantum systems~\cite{scovil1959three,doi:10.1063/1.461951,Harbola_2012,rahav-pra-2012,Paraoanu}, particle in a box~\cite{SU20182108,george-pra-2019}, and harmonic oscillators~\cite{PhysRevE.67.046105,lin2003optimization,andrea-pra-2016,kosloff2017quantum}, have already been employed in designing quantum heat engines.

In this article, we strive to construct a quantum heat engine whose working principle is exclusively based on   the quantum features of formation of quantized energy levels and quantum degeneracies~\cite{david-prl-2018,SU20182108,george-pra-2019}, owing to the finite boundary conditions~\cite{sakurai-2017}.
In this endeavour, we first propose a quantum Stirling heat engine based on an ensemble of quantum harmonic oscillators, where the degeneracy is generated by inserting a barrier in the middle of the harmonic oscillator. 
These degeneracies induce a lack of information, which can be converted to useful work using two  reservoirs  at different temperatures.
We provide a rigorous study of  work and efficiency in harmonic oscillator quantum Stirling heat engine (HO-QSHE), where the results show that   the efficiency can be maximized at a  specific  ratio of hot and cold reservoir temperatures,  and that this maximum  depends on the frequency of the harmonic oscillator; however, there is no such maximum for the extracted work. We also study the quantum Stirling heat engine based on an ensemble of particle-in-a-box  quantum systems. We discuss work and efficiency for  both symmetric and asymmetric insertion of  a single barrier and also the insertion of multiple barriers in particle in a box quantum Stirling heat engine (PIB-QSHE). The results reveal that both work and efficiency can be maximized at a certain specific  ratio of hot and cold temperatures, which depends on the length of the box.

The motivation behind choosing a quantum harmonic oscillator as the working medium of  the quantum heat engine is the fact that it is one of the most ubiquitous quantum systems, which models atoms in a lattice to quantum fields.
Since almost any generic potential can be approximated as a harmonic oscillator for small displacements near equilibrium~\cite{taylor-harmonic}, our proposed degeneracy assisted HO-QSHE may be realised practically.
Possible candidates include Bose-Einstein condensate, Josephson junction, and vibrational modes of solid~\cite{josephson-heat,Giovanni-ultra, Paraoanu_2001,weedbrook-rmp-2012}.
Similarly, quantum dots, wires, and wells are  promising candidates for the realisation of  the degeneracy assisted PIB-QSHE~\cite{sattler-2010}. 

We arrange the paper as follows. Sec.~\ref{sec:harmonicengine} describes  the quantum Stirling heat engine based on  harmonic oscillators. The work and efficiency for  it are analyzed in Sec.~\ref{sec:harmonicresults}, while the results for a quantum Stirling engine based on an ensemble of particle in a box quantum systems is discussed in Sec.~\ref{sec:boxresults}. In Sec.~\ref{sec:conclusion}, we discuss our results and future aspects of  this work. Appendix~\ref{appendix} recapitulates the particle in a box Stirling heat engine.

\section{Harmonic oscillator based quantum Stirling heat engine}
\label{sec:harmonicengine}
\begin{figure}[htbp]
		\includegraphics[scale=1]{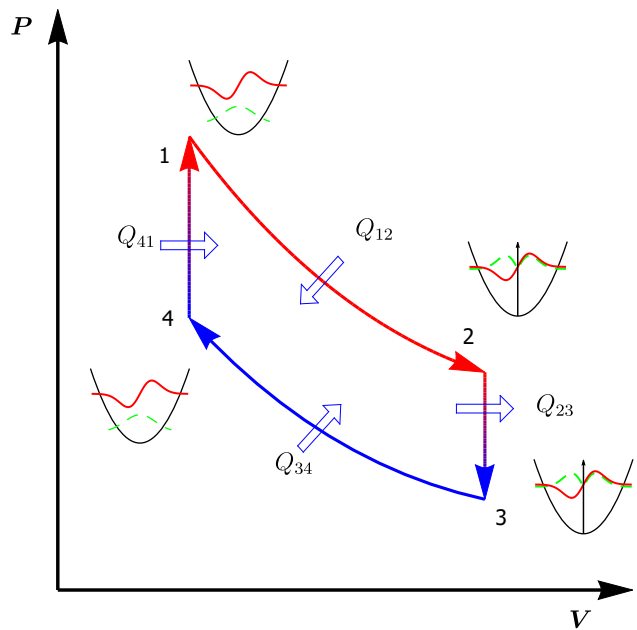} 
	\caption{  Schematic of the classical Stirling heat engine on a $PV$ diagram. 
The process $[(1) \rightarrow (2)]$ represents isothermal expansion of the system. The process $[(2) \rightarrow (3)]$ shows isochoric cooling of the engine when it is brought in contact with a reservoir at a lower temperature $T_c$. The process $[(3) \rightarrow (4)]$ represents isothermal compression and $[(4) \rightarrow (1)]$ is the isochoric heating when the engine is connected back to the heat reservoir at temperature $T_h$.
 The corresponding degeneracy assisted harmonic oscillator quantum Stirling heat engine is also depicted on the same diagram. 
  Here, a barrier is inserted at the centre of the harmonic oscillator during the process $[(1) \rightarrow (2)]$. Consequently, the even numbered energy levels are raised to the odd numbered energy levels and the final energy  spectrum is doubly degenerate. 
	The working medium during this process is constantly in equilibrium with the heat reservoir at temperature $T_h$. 
  During the process  $[(2) \rightarrow (3)]$ the system is connected to the reservoir at a lower temperature $T_c$ and as a result heat is released from the system.
 In the process $[(3) \rightarrow (4)]$, the barrier is removed quasi-statically from the harmonic oscillator while the Stirling heat engine is in equilibrium with a cold reservoir at temperature $T_c$. In the final process  of the cycle $[(4) \rightarrow (1)]$, the system is brought in contact with the heat reservoir at temperature $T_h$ and  heat is absorbed  by the system. 
	}
	\label{schematic} 
\end{figure}

In this section, we propose a quantum Stirling heat engine with the working medium as a harmonic oscillator. 
Analogous to the classical counterpart, a quantum Stirling cycle is a four-stroke closed cycle regenerative heat engine.  The schematic diagram of a classical Stirling heat cycle  has been shown in Fig.~\ref{schematic}.  The corresponding processes for the quantum Stirling heat engine are also superposed on the same diagram.

\par
The four steps involved  in a HO-QSHE are as follows.

\par
\noindent{\bf First step\,:}   In the first step of the quantum Stirling heat engine, a barrier is inserted at the centre of the  harmonic oscillator in a quasi-static manner, thus allowing the system to be in constant equilibrium with the heat bath at temperature $T_h$ through heat exchange between the two. 
In contrast, the first step in a classical heat engine is that of isothermal expansion, where a system  coupled to a   heat bath at temperature $T_h$ expands  isothermally (mechanical work is done by the system). 
It is worth mentioning that when we talk about the temperature of the system, we refer to the temperature of the ensemble of the working medium.
For a one dimensional quantum harmonic oscillator of frequency $\omega$  as the working medium, the $n^{th}$ energy level is given as
\begin{equation}
E_n = \bigg(n+\frac{1}{2}\bigg)\hbar\omega \quad \text{with}\, n=0,1,2,\dots,
\end{equation}
where  $\hbar$ is the reduced Planck constant.
This is the energy level spectra corresponding to the initial state for the first process of the HO-QSHE cycle  (stage $1$ in Fig.~\ref{schematic}). The partition function of the initial state $Z_{(1)}$ is given as
\begin{equation}
Z_{(1)} = \sum_{n=0}^{\infty} e^{-\frac{E_n}{k_B T_h}} = e^{\frac{-\hbar\omega}{2k_B T_h}} \sum_{n=0}^{\infty} e^{\frac{-n\hbar\omega}{k_B T_h}},
\end{equation}
 where $k_B$ is the Boltzmann constant.
The partition function is a geometric sum which can be readily evaluated as
\begin{equation}
Z_{(1)} = \frac{e^{\frac{-\hbar\omega}{2k_BT_h}}}{1-e^{\frac{-\hbar\omega}{k_BT_h}}} = \frac{1}{2 \sinh\bigg({\frac{\hbar\omega}{2k_BT_h}}\bigg)}.
\end{equation}
The isothermal process involves the quasi-static insertion of a barrier in  the centre of the harmonic oscillator.   For  this work, we assume that the centre of the potential is the origin of the coordinate system. This barrier is an infinite potential delta function ($p \hspace{2pt} \delta(0)$, $p\to\infty$), which introduces an additional constraint  of the probability amplitude of the wave function going to zero at the centre~\cite{Griffiths,boundary}.   A barrier inserted in this way does not alter the volume or the classical energy  of the system but affects the quantum mechanical energy wave functions. The barrier `splits' the wave functions into two exactly identical parts by introducing a node at  their midpoints.  Since the wave functions cannot vanish, the wave functions with even quantum numbers are `raised' in energy to the next odd numbered energy state with the introduction of the barrier at the origin. The energy states with odd quantum numbers, are now two-fold degenerate (stage $2$ in Fig.~\ref{schematic}). The new energy levels of this system are given as
 \begin{equation}
E_n = \bigg(n+\frac{1}{2}\bigg)\hbar\omega \quad \text{with}\, n=1,3,5,\dots,
\end{equation}
which could be alternatively written as
\begin{equation}
    E_n = \bigg( 2n + \frac{3}{2} \bigg)\hbar\omega \quad \text{with}\, n=0,1,2,\dots.
\end{equation}
 Various thermodynamics variables considered in this article are normalized to the number of particles in the ensemble, and therefore,  will be intensive. Therefore, the partition function can be written as
\begin{equation}
Z_{(2)} = 2e^{\frac{-3\hbar\omega}{2k_BT_h}}\sum_{n=0}^{\infty} 
 e^{-\frac{2n\hbar\omega}{k_BT_h}}=\frac{2 {e^{\frac{-3\hbar\omega}{2k_BT_h}}}} {\bigg(1-{e^{\frac{-2\hbar\omega}{k_BT_h}}}\bigg)}.
\end{equation}
The internal energy at stage (1)  of the Stirling cycle can be written in terms of the partition functions as
 $U_{(1)}= -\partial \ln Z_{(1)}/\partial \beta_h $, where $\beta_h$ is $1/k_B T_h$.
Therefore, the change in the internal energy from  stage $(1)$ to  stage $(2)$ can be expressed as
\begin{equation}\label{changeinternal}
\Delta U_{12} = U_{(2)}-U_{(1)}= -\frac{\partial}{\partial \beta_h} [\ln Z_{(2)} - \ln Z_{(1)}].
\end{equation}
Further, the thermodynamic entropy at stage (1) can be expressed in terms of the partition functions as
\begin{equation}\label{entropy}
    S_{(1)}= k_B \left(1- \beta_h \frac{\partial}{\partial \beta_h}\right) \ln Z_{(1)}.
\end{equation}
Therefore, the heat absorbed in the process   of taking the system from state $(1)$ to state $(2)$ is given as $Q_{12}= T_h( S_{(2)}- S_{(1)})$, which can be expressed as following using Eqs.~(\ref{entropy}) and (\ref{changeinternal}):
\begin{equation}
\begin{aligned}
Q_{12}&=  \left(\frac{1}{\beta_h}-  \frac{\partial}{\partial \beta_h}\right) [\ln Z_{(2)} - \ln Z_{(1)}],  \\
&=U_{(2)}-U_{(1)}+ k_{B}T_{h}\ln Z_{(2)}- k_{B}T_{h}\ln Z_{(1)}.
\end{aligned}
\end{equation}
We note that the change in internal energy  $(U_{(2)}-U_{(1)})$ is not zero  since the energy level spectra of the harmonic oscillator changes when the barrier is inserted. This is in contrast with the classical Stirling heat engine, where  the internal energy remains constant during an isothermal process. This is because the energy level spectra of the working medium  is  the same throughout the cycle. 

The work done in the process is given by first law of thermodynamics as
\begin{equation}
W_{12}= Q_{12} - \Delta U_{12} =k_{B}T_{h}\ln Z_{(2)}- k_{B}T_{h}\ln Z_{(1)}.
\end{equation}
The work in this step is done by the system and hence is expected to be negative. 

\par
\noindent{\bf Second step\,:}
 In this step, we connect the system  to a thermal bath at temperature $T_c<T_h$ after disconnecting it from the thermal bath at temperature $T_h$. Consequently, the temperature of the system falls down from $T_h$ to $T_c$. This process for a classical Stirling heat engine is termed isochoric cooling, where the volume remains constant, and thus no mechanical work is done. Therefore, this process causes the system to lose heat.

The partition function of the initial state for this step is $Z_{(2)}$. The partition function of the final state is exactly similar to $Z_{(2)}$,  since the energy spectra of the system remains the same, with the exception  that here the temperature is $T_c$ instead of $T_h$ . The partition function is given as 
\begin{equation}
Z_{(3)} = 2e^{\frac{-3\hbar\omega}{2k_BT_c}}\sum_{n=0}^{\infty}  e^{-\frac{2n\hbar\omega}{k_BT_c}}= \frac{2 {e^{\frac{-3\hbar\omega}{2k_BT_c}}}} {\bigg(1-{e^{\frac{-2\hbar\omega}{k_BT_c}}}\bigg)}.
\end{equation}

The amount of heat lost in this step can be calculated in terms of the partition function. Since no mechanical work is done ($W_{23} = 0$), the heat lost is equal to the change in internal energy:
\begin{equation}
Q_{23} = U_{(3)} - U_{(2)} = -\frac{\partial \ln Z_{(3)}}{\partial \beta_c} + \frac{\partial \ln Z_{(2)}}{\partial \beta_h}.
\end{equation}

\par
\noindent{\bf Third step\,:} 
In the next step, the barrier is  removed quasi-statically such that the system is in thermal equilibrium with the cold bath at temperature $T_c$ at all times. At the end of the process, the energy spectra become identical to that of the initial state $(1)$. 
Again the volume remains constant  for the quantum Stirling heat engine. In contrast, the corresponding step in a classical heat engine is that of isothermal compression, where a system  coupled to a cold bath at temperature $T_c$ undergoes a compression.
The partition function of the final state is same as that of $Z_{(1)}$, but with temperature $T_c$  instead of $T_h$:
\begin{equation}
   Z_{(4)} =  e^{\frac{-\hbar\omega}{2k_B T_c}} \sum_{n=0}^{\infty} e^{\frac{-n\hbar\omega}{k_B T_c}} = \frac{e^{\frac{-\hbar\omega}{2k_BT_c}}} {\bigg(1-{e^{\frac{-\hbar\omega}{k_BT_c}}}\bigg)}.
\end{equation}

The heat provided to the system at the end of the process is obtained in terms of the partition functions as
\begin{equation}
Q_{34}= U_{(4)}-U_{(3)}+ k_{B}T_{c}\ln Z_{(4)}- k_{B}T_{c}\ln Z_{(3)}.
\end{equation}
The isothermal work done is obtained from the first law of thermodynamics as
\begin{equation}
W_{34} = Q_{34} - \Delta U_{34} = k_{B}T_{c}\ln Z_{(4)}- k_{B}T_{c}\ln Z_{(3)},
\end{equation}
where $\Delta U_{34} = U_{(4)} - U_{(3)}$ is the change in internal energy in the process.
\par
\noindent{\bf Fourth step\,:} In the final step, the system is detached from the cold bath at temperature $T_c$ and connected to the hot bath at temperature $T_h$. This raises the temperature of the system from $T_c$ to $T_h$.
The corresponding process for a classical Stirling heat engine is called isochoric heating, where the volume does not change, and therefore no mechanical work is done. Thus, the system loses heat during this process.
Thus at the end of the process, the system returns to the initial state $(1)$ (the ensemble of harmonic oscillators at a temperature $T_h$). Since no work is done in this step, the heat absorbed by the system is equal to the change in internal energy:
\begin{equation}
Q_{41}= -\frac{\partial \ln Z_{(1)}}{\partial \beta_h} + \frac{\partial \ln Z_{(4)}}{\partial \beta_c}.
\end{equation}

The four thermodynamic processes described above  form one complete cycle of the quantum Stirling heat engine. The total work done by the system in one complete cycle  of the HO-QSHE is given as
\begin{eqnarray}\label{workho}
W_{\textrm{net}} &=& W_{12} + W_{34},\nonumber \\
&=& k_{B}T_{h}\ln \frac{Z_{(2)}}{Z_{(1)}}+k_{B}T_{c}\ln \frac{Z_{(4)}}{Z_{(3)}}. 
\end{eqnarray}

Further, since the system absorbs heat only during the first isothermal step $[(1) \rightarrow (2)]$ and the final isochoric step $[(4) \rightarrow (1)]$, the total heat absorbed during one complete cycle of the HO-QSHE can be written as
\begin{equation}
Q_{\textrm{in}} = Q_{12} + Q_{41}.
\end{equation}
The efficiency $\eta$ for the cycle is given in terms of the work done and the heat intake as
\begin{equation}
\label{efficiencyho}
\eta = \frac{W_{\textrm{net}}}{Q_{\textrm{in}}}  = 1+\frac{Q_{23}+Q_{34}}{Q_{12}+Q_{41}}.
\end{equation}
The expression for the efficiency can be obtained from the partition functions and their derivatives. 

\section{Analysis of work and efficiency}
\label{sec:horesults}
In this section, we first analyse the work and efficiency for  HO-QSHE and then move on to PIB-QSHE. 
\subsection{Work and efficiency  of the harmonic oscillator quantum Stirling heat engine}\label{sec:harmonicresults}
We first explore the work and efficiency of the HO-QSHE with respect to the frequency  $\omega$ of the harmonic oscillator for different values of $T_h/T_c$. For convenience, we  introduce dimensionless units for the work extracted and the frequency by re-scaling them as $W/k_B T_c$ and as $\omega \hbar/k_B T_c$ respectively.

\begin{figure}[htbp]
	\includegraphics[scale=1]{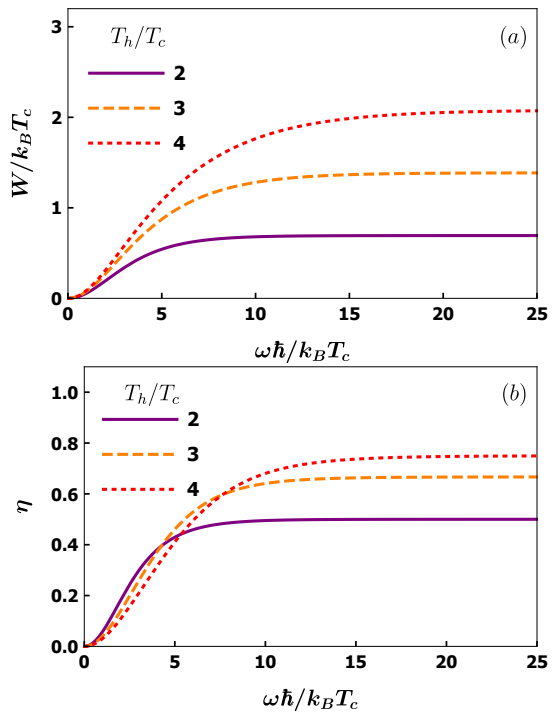} 
	\caption{(a) Plot of  work $W/k_B T_c$ as a function of  frequency $\omega \hbar/k_B T_c$ for different values of $T_h/T_c$. (b) Plot of efficiency $\eta$  as a function of  frequency $\omega \hbar/k_B T_c$ for different values of $T_h/T_c$. }
	\label{fig:harmonicfreq} 
\end{figure}
We define the harmonic oscillator of frequency $\omega$ to be in the low temperature limit, when the following condition is satisfied:
\begin{equation}
    \hbar \omega >> k_B T_h \quad \text{with} \,\,T_h>T_c.
\end{equation}
 This condition is satisfied for  large values of $\omega$ as well as very small values of $T_h$. 

As shown in Fig.~\ref{fig:harmonicfreq}(a), the work  increases  monotonically from zero  as a function of the frequency of the harmonic oscillator and  asymptotically approaches the high frequency limit, which can be derived using Eq.~(\ref{workho})  to be $(T_h/T_c-1)\ln 2$. This result can be numerically verified from the plot. 
Furthermore, as we increase the value of $T_h/T_c$, the amount of extracted work  increases.

Similarly,  the efficiency also starts to rise from zero as  the frequency increases and becomes asymptotic in the high frequency limit as shown in Fig.~\ref{fig:harmonicfreq}(b). We can derive this asymptotic efficiency value  using Eq.~(\ref{efficiencyho}) as $(1-T_c/T_h)$ which can also be verified from the figure.
 Interestingly, this is the efficiency of a classical Carnot cycle. Thus, in the high frequency or low temperature limit, the efficiency of  a HO-QSHE approaches the Carnot  efficiency.

We note that efficiency curves for different values of $T_h/T_c$ cross over each other as the frequency is varied.

\begin{figure}[htbp]
	\includegraphics[scale=1]{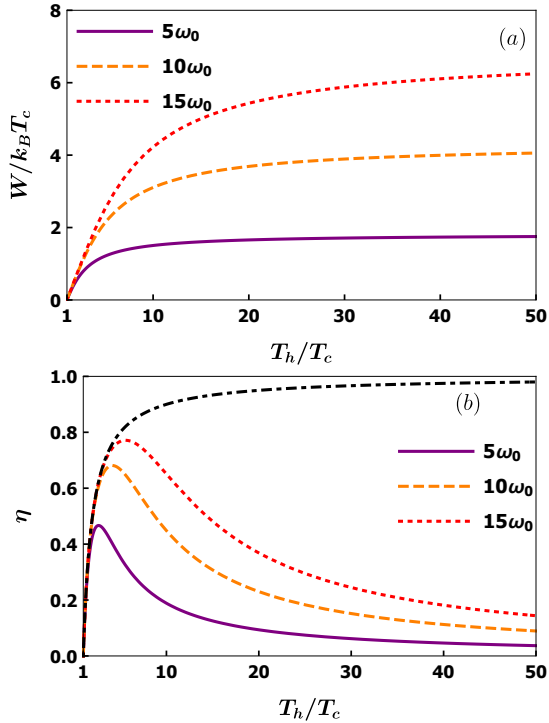} 
	\caption{(a) Plot of  work $W/k_B T_c$ as a function of $T_h/T_c$ for different frequencies of the harmonic oscillator. (b) Plot of efficiency $\eta$ as a function of $T_h/T_c$ for different  frequencies of the harmonic oscillator. The values considered  for the harmonic oscillator frequency are $\omega=5\,\omega_0$, $\omega=10\,\omega_0$, and $\omega=15\,\omega_0$, where $\omega_0= k_B T_c/\hbar$.  The black dot dashed line represents the Carnot  limit.} 
	\label{fig:harmonictemp} 
\end{figure}

\begin{figure}[htbp]
	\includegraphics[scale=1]{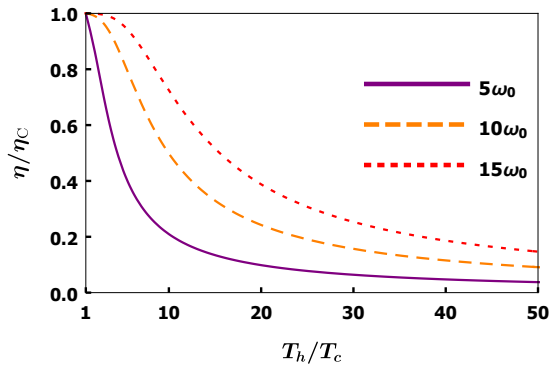} 
	\caption{ Plot of  normalized efficiency $\eta/\eta_{\text{C}}$  as a function of $T_h/T_c$ for HO-QSHE. Here $\eta_{\text{C}}$ denotes Carnot limit. }
	\label{fig:effharmonic} 
\end{figure}

To analyse the explicit temperature dependence, we plot work and efficiency as a function of $T_h/T_c$  for different frequencies of the harmonic oscillator in Fig.~\ref{fig:harmonictemp}. Here, the frequencies of the harmonic oscillator  have been set to be  integral multiples of $\omega_0 \; (\omega=m\omega_0)$, where $\omega_0=k_B T_c/\hbar$. The results show that the extracted work starts to increase from zero  with increase in the value of $T_h/T_c$ and becomes asymptotic in the high $T_h/T_c$ limit. The asymptotic value of the net work extracted can be computed using  Eq.~(\ref{workho}) as
\begin{equation}
    \lim\limits_{ T_h/T_c \to \infty} W_{\textrm{net}}=k_B T_c\left(\ln \left[\frac{1+e^m}{2}\right]-\frac{m}{2}\right).
\end{equation} 
Therefore, the scaled work is given as  $\ln [(1+e^m)/2]-m/2$, which can be numerically verified from the figure.
 On the other hand,  the efficiency as a function of $T_h/T_c$ attains a maximum after a steep ascent and then gradually levels off to zero in the asymptotic limit. In contrast, the efficiency of a classical heat engine increases monotonically as a function of $T_h/T_c$.  We further notice from Fig.~\ref{fig:harmonictemp}(b) that the efficiency approaches Carnot limit for small $T_h/T_c$. This fact can be verified from Fig.~\ref{fig:effharmonic}, where we have plotted  efficiency normalized by Carnot limit  as a function of $T_h/T_c$.

\begin{figure}[htbp]
	\includegraphics[scale=1]{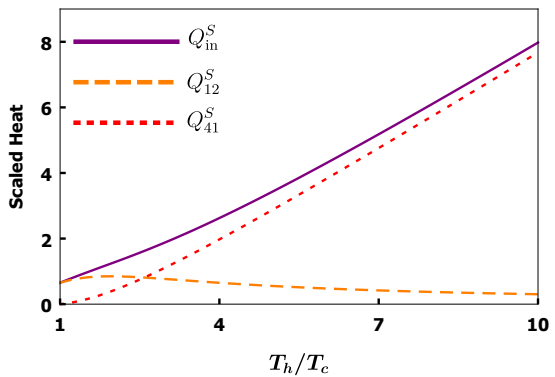} 
	\caption{ Plot of  heat ($Q/k_B T_c$) as a function of $T_h/T_c$. The superscript $`S\textrm'$ represents the corresponding scaled quantities. We have set the frequency of the harmonic oscillator to be $\omega = 5\, \omega_0$.  }
	\label{fig:qscaled} 
\end{figure}

Furthermore, the monotonic decrease of the efficiency in Fig.~\ref{fig:harmonictemp}(b) after the maxima and consequently the asymptotic  decay to zero in the high $T_h/T_c$ limit can be attributed to the fact that as the 
 ratio $T_h/T_c$ increases, more heat needs to be 
 provided to the system to raise  its temperature from $T_c$ to $T_h$ in the final step [$(4) \rightarrow (1)$]  of the Stirling cycle. This increases the value of total heat input $Q_\textrm{in}$  to the system. As can be seen from Fig.~\ref{fig:qscaled}, the heat taken  in during the isothermal process $Q_{12}$ also contributes to $Q_\textrm{in}$, but its effect is small  since there is no temperature change during  this process.  
 
Thus, because of the $Q_{41}$ term, $Q_\textrm{in}$ dominates over  $W_\textrm{{net}}$,  consequently diminishing the efficiency of the cycle as per Eq.~(\ref{efficiencyho}), and thus the efficiency approaches  zero in the high $T_h/T_c$  limit.

It should be noted that the curves  plotted as a function of $T_h/T_c$ start from $T_h/T_c=1$, \ie, when both the reservoirs are at the same temperature, and hence no work is extracted.
We have  numerically computed  the values of $T_h/T_c$ corresponding to the maximum efficiency for different harmonic oscillator frequencies  and compared it with the Carnot efficiency at the corresponding values of $T_h/T_c$  in Table~\ref{table1}.  The results  show that  although increasing the frequency initially leads to increase in the efficiency of the HO-QSHE, the rate of increase slows down and eventually saturates. Further,  the efficiency for the HO-QSHE is bounded from above by the Carnot efficiency. The maximum efficiency asymptotically approaches unity as the frequency of the harmonic oscillator $\omega \rightarrow \infty$. Furthermore, the unit efficiency is achieved for $T_h/T_c \rightarrow \infty$ as is for a Carnot engine. This behavior is reinforced by Fig.~\ref{fig:harmonictemp}.  
\begin{table}[ht!]
	\caption{\label{table1}
		Comparison of the numerical results for maximum efficiency of the HO-QSHE ($\eta_{\text{max}}$) and Carnot efficiency ($\eta_{\text{C}}$) }
		\begin{tabular}{ p{2.5cm} p{2cm} p{2cm} p{1.5cm}}
			\hline \hline
			Frequency $\omega$ & $T_h/T_c$ & $\eta_{\text{max}}$ & $\eta_{\text{C}}$ \\
			\hline 
		$5\,\omega_0$&    2.66&0.47 & 0.62 \\ \hline
		$10\,\omega_0$&4.15&0.68& 0.76 \\	\hline
		$15\,\omega_0$&5.58&0.77 & 0.82\\ \hline
		$50\,\omega_0$&14.57&0.92  & 0.93 \\ \hline
		$150\,\omega_0$&36.93&0.97  &  0.97\\ \hline
		$350\,\omega_0$&77.14&0.98  & 0.99\\ \hline
			\hline
		\end{tabular}
	\end{table}

The cross over of the efficiency curves in Fig.~\ref{fig:harmonicfreq}(b) is a consequence of the existence of maxima  in Fig.~\ref{fig:harmonictemp}(b). 
 We try to explain our point by comparing the value of efficiency at two different frequencies for different values of $T_h/T_c$.
For instance, the efficiency  curve for $\omega =5\, \omega_0$, which has a maximum at $T_h/T_c=2.66$, takes the value $\eta = 0.43$, $0.46$, and $0.41$ at $T_h/T_c=2$, $3$, and $4$, respectively. Hence, at $\omega =5\, \omega_0$, the efficiency is maximum for $T_h/T_c=3$ and minimum for $T_h/T_c=4$, which can be confirmed from  Fig.~\ref{fig:harmonicfreq}(b).  Similarly, the efficiency  curve for $\omega =10 \,\omega_0$ takes the value $\eta = 0.49$, $0.64$, and $0.68$ at $T_h/T_c=2$, $3$, and $4$, respectively. 
Hence,  the efficiency is maximum for $T_h/T_c=4$ and minimum for $T_h/T_c=2$ at $\omega =10\, \omega_0$.
These  changes in the ordering of the numerical values of  efficiency for different values of $T_h/T_c$ causes a crossover of the efficiency curves.
	
	 It is worth noting that the limit when the frequency of the harmonic oscillator approaches zero corresponds to the classical limit. Since the gap between consecutive energy levels goes to zero as $\omega \rightarrow 0$, there will be a continuum of energy levels. Since the particle is like a free particle in this limit, the work needed to insert or remove the barrier becomes zero. This feature in the context of particle in a box quantum Szilard engine has been explored in Ref.~\cite{classical-ap-2012}.
\subsection{Work and efficiency of particle in a box quantum Stirling heat engine}
\label{sec:boxresults}
In this section, we consider PIB-QSHE, which was proposed in~\cite{george-pra-2019}.  While a preliminary numerical analysis of PIB-QSHE can be found in~\cite{george-pra-2019}, we study it in quite detail and provide important insights into the behaviour of net work and efficiency of the PIB-QSHE.
In this regard, we first analyse  the net work and efficiency when a single barrier is inserted symmetrically, \ie, in the middle of the box. We then consider  the symmetric insertion of multiple barriers and finally move on to the case, where a single barrier is inserted asymmetrically into the box. We have described the  respective steps in the Stirling cycle for these different cases in Appendix~\ref{appendix}.
\subsubsection{Symmetric insertion of a single barrier}\label{singlebarrier}
\begin{figure}[htbp]
	\includegraphics[scale=1]{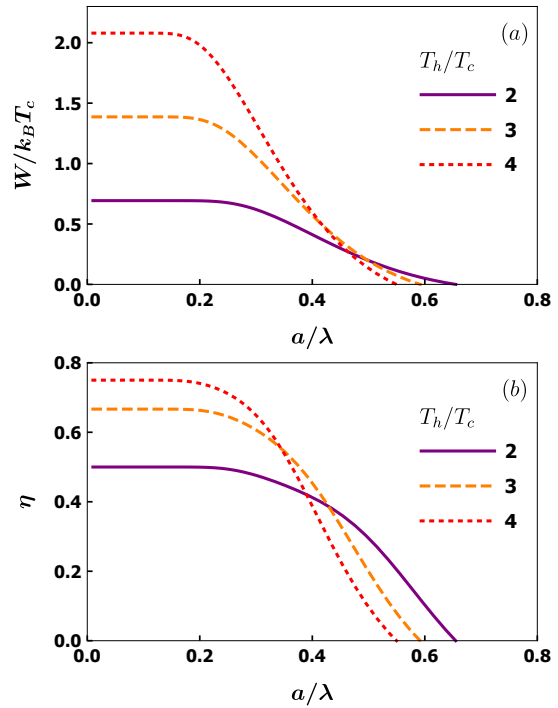} 
	\caption{(a) Plot of  work $W/k_B T_c$ as a function of  length of the box $a/ \lambda$ for different values of $T_h/T_c$. (b) Plot of efficiency $\eta$ as a function of  length of the box $a/\lambda$ for different values of $T_h/T_c$.	}
	\label{fig:singlelength} 
\end{figure}
We consider the scenario where a  barrier is inserted at the centre of a box of length $2a$. We have described the different stages of PIB-QSHE in Appendix~\ref{appendixA}.
We first analyse the work and efficiency of PIB-QSHE with respect to the length of the box. We re-scale the work as $W/k_B T_c$ and the length of the box as $a/\lambda$, where $\lambda= h/ \sqrt{2 m k_B T_c}$ is the thermal de Broglie wavelength.

We  define the box of length $2a$ to be in the low temperature limit, when the condition
\begin{equation}
    \frac{\pi^2 \hbar^2}{2 m (2a)^2}>>k_B T_h, \quad \text{with} \,\,T_h>T_c,
\end{equation}
holds. We also note that  the above condition is satisfied for small lengths of the box as well as  for small temperatures $T_h$.

The low-temperature limit of the scaled work~(\ref{pib:workeqn}) turns out to be $(T_h/T_c-1) \ln 2$. This numerical value can also be confirmed  for small  $`a\textrm'$ values from Fig.~\ref{fig:singlelength}(a). On further increasing the value of the length of the box $a/\lambda$, the work starts to decrease and eventually becomes zero for a particular value of $a/\lambda$. For $T_h/T_c =2$, the value of $a/\lambda $ for which the work becomes zero turns out to be $0.65$. 
Similarly, the low temperature limit of the efficiency expression~(\ref{pib:effeqn}) turns out to be $1-T_c/T_h$, which can  also be numerically verified from Fig.~\ref{fig:singlelength}(b).
 The efficiency also starts to drop with an increase in the value of   $a/\lambda$ and becomes zero at the same value where work becomes zero.   
 
  As we increase the value of $T_h/T_c$  in Fig.~\ref{fig:singlelength}, the  work and efficiency  corresponding to a particular value of $a/\lambda$ increases till the low temperature limit is satisfied. This is also seen from the  low temperature limit expressions for  work and efficiency.
 As we move away from the low temperature limit, crossover between work as well as efficiency curves is seen. Furthermore, it can be seen that with an increase in $T_h/T_c$, the work and efficiency reach zero for smaller values of $a/\lambda$.

\begin{figure}[htbp]
	\includegraphics[scale=1]{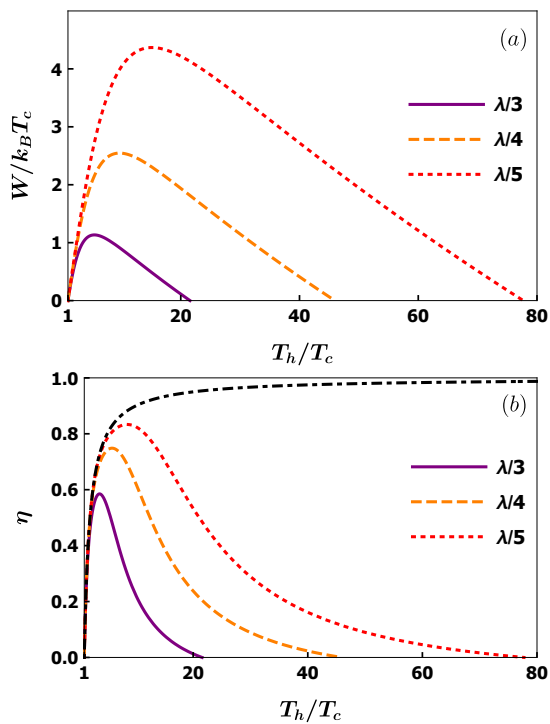} 
	\caption{(a) Plot of work $W/k_B T_c$ as a function of $T_h/T_c$ for different length of the box. (b) Plot of efficiency $\eta$ as a function of $T_h/T_c$ for different length of the box. The values considered are $a=\lambda/3$, $\lambda/4$, and $\lambda/5$, where $\lambda= h/ \sqrt{2 m k_B T_c}$ is the thermal de Broglie wavelength.  The black dot dashed line represents the Carnot  limit.	}
	\label{fig:singletemp} 
\end{figure}
We now study the dependence of work and efficiency on the temperature of the system. We take the length of the box as a fraction of the de Broglie wavelength and plot work and efficiency  with respect to $T_h/T_c$ in Fig.~\ref{fig:singletemp}. The results show that as the length of the box decreases, work and efficiency increase. We already saw this property in Fig.~\ref{fig:singlelength}.

Furthermore, both work and efficiency attain a maximum at a certain specific value of $T_h/T_c$. The maxima of these plots is the reason for the crossover seen in Fig.~\ref{fig:singlelength}. From Fig.~\ref{fig:singletemp}(a), we see that the  work becomes zero at larger values of $T_h/T_c$ as the  length of the box decreases.  So, if we plot the  work as a function of  length, Fig.~\ref{fig:singlelength}(a), the plot for higher $T_h/T_c$ will go to zero earlier, crossing over the other curves on its way. The crossover for the efficiency curve in Fig.~\ref{fig:singlelength}(b) is due to the same reason.

\begin{figure}[htbp]
 	\includegraphics[scale=1]{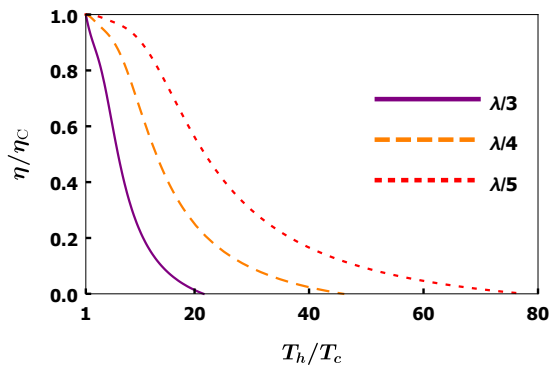} 
 	\caption{ Plot of  normalized efficiency $\eta/\eta_{\text{C}}$  as a function of $T_h/T_c$ for PIB-QSHE.  }
 	\label{fig:effbox} 
 \end{figure}

 From Fig.~\ref{fig:singletemp}(b), we also see that the efficiency approaches Carnot limit for small $T_h/T_c$. This can be explicitly seen in Fig.~\ref{fig:effbox},
 where we have plotted  normalized efficiency $\eta/\eta_{\text{C}}$  as a function of $T_h/T_c$.  
  
   We now compare numerically the maximum efficiency of the PIB-QSHE with the Carnot efficiency in Table~\ref{table2}.  We note that   this behaviour  is similar to that of the HO-QSHE.  Although the maximum efficiency  increases as the length of the box decreases, it always lies below the Carnot efficiency. Thus, the efficiency of the PIB-QSHE is also bounded by the Carnot efficiency  from above. In the limit of $a \rightarrow 0$, the maxima of the efficiency asymptotically approach unity. 
   \begin{table}[ht!]
	\caption{\label{table2}
		Comparison of the numerical results for maximum efficiency of the PIB-QSHE ($\eta_{\text{max}}$) and Carnot efficiency ($\eta_{\text{C}}$) }
		\begin{tabular}{ p{2.5cm} p{2cm} p{2cm} p{1.5cm}}
			\hline \hline
			Length $a$ & $T_h/T_c$ & $\eta_{\text{max}}$ & $\eta_{\text{C}}$ \\
			\hline 
		$\lambda/3$&  3.669  &0.585 & 0.727 \\ \hline
		$\lambda/4$& 5.850 & 0.749 & 0.829 \\ \hline
		$\lambda/5$& 8.488 & 0.833 & 0.882 \\ \hline
		$\lambda/10$& 28.067 & 0.954 & 0.964 \\ \hline
		$\lambda/20$& 95.988 & 0.987 & 0.990 \\ \hline
			\hline
		\end{tabular}
	\end{table}


 We now consider the limit of the length of box becoming large, \ie, $a \rightarrow \infty$. The gap between two energy levels approaches zero as  $a \rightarrow \infty$, and we obtain a continuum of energy levels. Therefore, the particle behaves as a free particle and the work needed to insert or remove the barrier becomes zero.

 
\subsubsection{Symmetric insertion of multiple barriers} \label{multiplebarrier}

\begin{figure}[htbp]
	\includegraphics[scale=1]{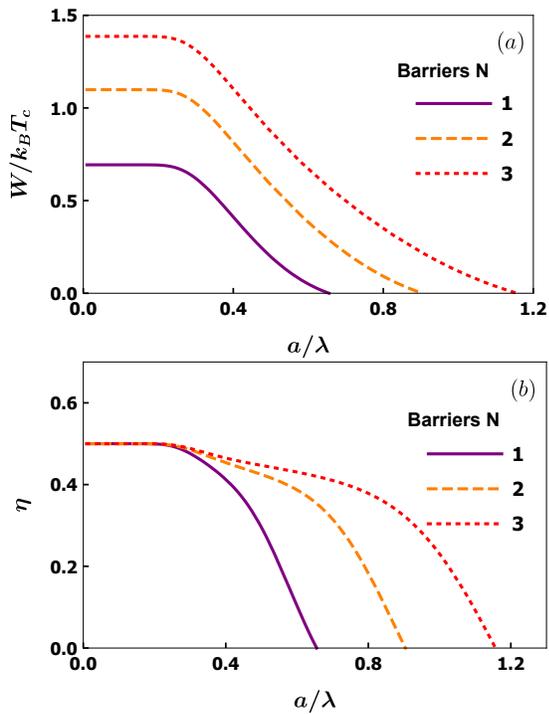} 
	\caption{(a) Plot of  work $W/k_B T_c$ as a function of length of the box $a/\lambda$ for different number of barriers. (b) Plot of efficiency $\eta$ as a function of  length of the box $a/\lambda$ for different number of barriers. The value of $T_h/T_c$ is taken to be $2$. }
	\label{fig:multiplelength} 
\end{figure}

This section studies work and efficiency in the scenario where we insert $N$  barriers symmetrically into the box. The details of the Stirling cycle  for this process has been provided in Appendix~\ref{appendixC}.
We analyse the  work and efficiency for multiple barrier insertions as functions of the length of the box and $T_h/T_c$, similar to that in Sec.~\ref{singlebarrier},  and we use the partition functions~Eq.~(\ref{multiplep1}) to Eq.~(\ref{multiplep4}).

In the low temperature limit, the  work for multiple barrier insertion case attains a value of $(T_h/T_c-1) \ln (N+1)$, which can be numerically verified from Fig.~\ref{fig:multiplelength}(a). The work, therefore, increases with  an increase in  the number of barriers.

 We explain the work extracted in the low temperature limit due to the degeneracy of the lowest energy level as it is  the only energy level which is accessible in   this limit. We note that $N+1$ is the degeneracy of the energy levels of the box  after symmetric insertion of $N$ barriers. Therefore, 
 the factor $\ln (N+1)$ arising in the work can be attributed to the degeneracy of the ground state energy level.
 As pointed out earlier, it is the lack of information  due to this degeneracy that is converted to work in our degeneracy assisted quantum heat engine. Hence in the low temperature limit, the work  is entropic in the sense that it is proportional to the Boltzmann entropy given as $S = k_B \ln (N+1)$. A direct verification of this claim can be stated. If the barrier is inserted in a manner such that no degeneracy is attained, we would expect the low temperature limit of the work obtained to be zero. This is one of the motivations behind the analysis of asymmetric barrier insertion in the next section.

In the low temperature limit, the efficiency turns out to be  $1-T_c/T_h$, which can be numerically verified from Fig.~\ref{fig:multiplelength}(b). Thus, in the low temperature limit, efficiency is independent of the number of barriers. However, in the high temperature limit, efficiency does depend on the number of barriers.
Further, as we increase the number of barriers, work and efficiency become zero at larger values of $a/\lambda$.

\begin{figure}[htbp]
	\includegraphics[scale=1]{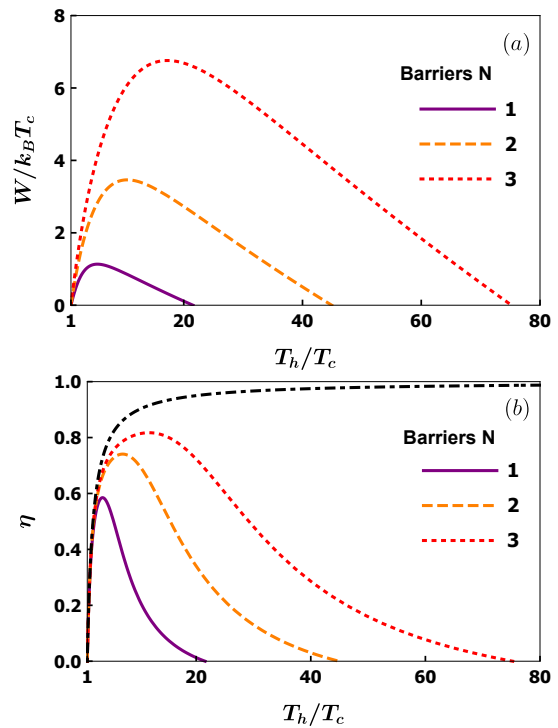} 
	\caption{(a) Plot of  work $W/k_B T_c$ as a function of $T_h/T_c$ for different number of barriers. (b) Plot of efficiency $\eta$ as a function of $T_h/T_c$ for different number of barriers. The half-length of the box is taken to be $a=\lambda/3$, where $\lambda= h/ \sqrt{2 m k_B T_c}$ is the thermal de Broglie wavelength.  The black dot dashed line represents the Carnot  limit.}
	\label{fig:multipletemp} 
\end{figure}

We also plot work and efficiency with respect to $T_h/T_c$ in Fig.~\ref{fig:multipletemp} for different number of barriers. We can see that as $T_h/T_c$ increases, both work and efficiency attain a maximum and then gradually come down to zero for any fixed number of barriers.  We also observe that as the number of barriers increases, the maximum value of both work and efficiency increases. Further, with an increase in the number of barriers, the maximum as well as zero shifts to higher values of $T_h/T_c$.

\subsubsection{Asymmetric insertion of a single barrier}\label{asymmetricbarrier}
\begin{figure}[htbp]
	\includegraphics[scale=1]{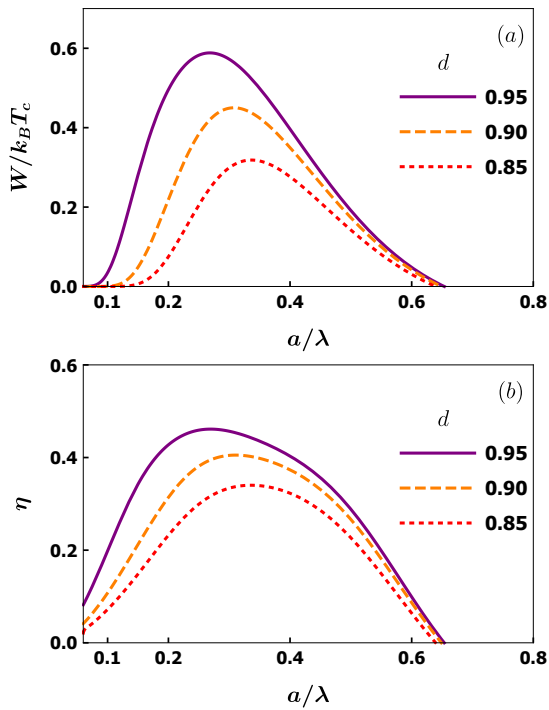} 
	\caption{(a) Plot of  work $W/k_B T_c$ as a function of length of the box $a/ \lambda$ for different ratios $d$ of the box  lengths. (b) Plot of efficiency $\eta$ as a function of length of the box $a/\lambda$ for different ratios of the box lengths. The value of $T_h/T_c$ is taken to be $2$. }
	\label{fig:asymlength} 
\end{figure}

\begin{figure}[htbp]
	\includegraphics[scale=1]{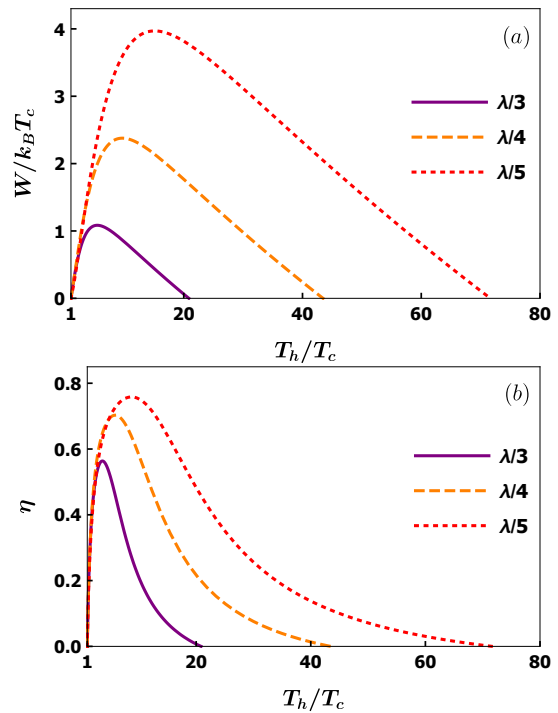} 
	\caption{(a) Plot of work $W/k_B T_c$ as a function of $T_h/T_c$ for different length of the box. (b) Plot of efficiency $\eta$ as a function of $T_h/T_c$ for different length of the box. The values considered are $a=\lambda/3$, $\lambda/4$, and $\lambda/5$, where $\lambda= h/ \sqrt{2 m k_B T_c}$ is the thermal de Broglie wavelength. In both the plots, we have taken $d=0.95$.	}
	\label{fig:asymtemp} 
\end{figure}

In this section, we examine  the work and efficiency for  an asymmetric insertion of a barrier  in the box, as  functions of the length of the box and $T_h/T_c$, and we use the partition functions~Eq.~(\ref{asymp1}) to Eq.~(\ref{asymp4}). This process divides the box into two parts of lengths $x$ and $y$ with $x+y=2a$. We provide the details of the Stirling cycle for this case  in Appendix~\ref{appendixB}.
 We first study the work $W/k_BT_c$ as a function of the length of the box  $a/\lambda$ for different values of  $`d\textrm'$, which we define to be the ratio of the lengths of the  two box parts, \ie,  $d=x/y$. The results are shown in Fig.~\ref{fig:asymlength}. The plots reveal that both work and efficiency attain a maximum at a certain value of $a/\lambda$. This result is completely different from the symmetric barrier insertion case Fig.~\ref{fig:singlelength}.  Interestingly, the work in the low temperature limit goes to zero as predicted from the $\ln (N+1) $ dependence, (where $N+1$ is the degeneracy of the lowest energy level). The rationale behind this behaviour would be that in the low temperature limit, only the lowest energy level is accessible, which is non-degenerate in the case of asymmetric barrier insertion, and consequently the work reduces to zero.
 
 We now move on to study the work and efficiency dependence on the temperature for different  half-lengths of the box $a=\lambda/3$, $\lambda/4$, and $\lambda/5$, where $\lambda= h/ \sqrt{2 m k_B T_c}$ is the thermal de Broglie wavelength. The results are shown in Fig.~\ref{fig:asymtemp}. We have set $d=0.95$ in both the plots.    We observe that both work and efficiency attain a maximum at a certain specific value of $T_h/T_c$. This result is similar compared to the symmetric barrier insertion case Fig.~\ref{fig:singletemp}; however, the magnitude of work and efficiency decreases in the asymmetric  insertion case compared to the symmetric one.
 
 \section{Concluding remarks} 
\label{sec:conclusion} 
In this paper, we proposed a degeneracy assisted HO-QSHE in this work and  analysed the work and efficiency as  functions of the frequency of the harmonic oscillator and the ratio $T_h/T_c$ of temperatures of the  hot and cold thermal baths.
We also examined PIB-QSHE in full detail, which was proposed in~\cite{george-pra-2019}.  We note that  the energy levels are inhomogeneously scaled upon symmetric insertion of a single barrier in harmonic oscillator and particle in a box~\cite{david-prl-2018}.  We showed that efficiency is maximized at  certain  temperature ratios for both HO-QSHE and PIB-QSHE. However,  the work extracted can only be maximized for PIB-QSHE at a certain temperature ratio.
It remains an open problem to find out the reason behind this contrasting behaviour.
In the low temperature limit, efficiency of both the HO-QSHE and PIB-QSHE approach the Carnot efficiency.

 We would like to point out one important distinction between a harmonic oscillator and particle in a box. The two systems differ in the  way their specific heat capacity  behaves as a function of temperature.  This is  essentially a consequence of the  difference in the structures of the energy levels in a harmonic oscillator and particle in a box. For a one dimensional harmonic oscillator, the heat capacity per particle increases  monotonically with temperature  and asymptotically  reaches $k_B$ in the infinite temperature limit. On the other hand, for a one dimensional particle in a box, the heat capacity per particle starts to increase with temperature and  attains a maximum value of approximately $9k_B/16$ and  in the high-temperature limit approaches $k_B/2$ from the above~\cite{box-ajp-1962}. 
A careful analysis of these facts will provide more insights into the working principle of degeneracy assisted quantum heat engines and might help in resolving the aforementioned problem.

 Quantum heat engines have already been realised on several different systems, for instance, quantum dots~\cite{martin-nature-2018}, cold bosonic ions~\cite{fialko2012isolated}, optomechanical systems~\cite{zhang2014optomech}, and liquid NMR based platforms~\cite{batalhao2014experimental}.   The degeneracy assisted quantum heat engine proposed in this work may be practically realised in the near future and our theoretical analysis would be  useful in  operating the quantum heat engine at optimal conditions.

 As we have mentioned earlier that efficiency of quantum heat engines can go beyond the Carnot limit, it would be interesting to see  this effect for degeneracy assisted quantum heat engines  by considering different type of baths. Another interesting direction is to construct a quantum engine based on entangled states of the harmonic oscillator (an infinite-dimensional system). Entangled quantum heat engines based on two qubit systems~\cite{ting-pra-2007,george-pra-2011,yong-physica-2020} (a finite-dimensional system) have already been proposed.

\section*{Acknowledgement}
A.K. and C.K. thank Narayansami Sathyamurthy for initial discussions on the subject.
 All the authors thank George Thomas, Narayansami Sathyamurthy, and Upendra Harbola 
 for their invaluable comments on the final version of the draft. 
 We dedicate this work to Prof. N. Sathyamurthy's forthcoming 70th birthday.  
 C.K. acknowledges the financial
support from {\bf DST/ICPS/QuST/Theme-1/2019/General} Project
number {\sf Q-68}.

\appendix
\section{Particle in a box based quantum Stirling heat engine: background material}\label{appendix}

In the Appendix, we discuss quantum Stirling heat engine with a particle in a box as the working medium (PIB-QSHE)~\cite{george-pra-2019}. We consider three different scenarios in details: (i) symmetric insertion of a single barrier, (ii) symmetric insertion of multiple barriers, and (iii) asymmetric insertion of a single barrier.

\subsection{Symmetric insertion of single barrier}\label{appendixA}
We first consider symmetric insertion of single barrier in a box of length $2a$.
\par
\noindent{\bf First step\,:}   
In the first step, we insert a barrier in the middle of the box, which is coupled to a thermal bath at temperature $T_h$. The insertion of the barrier is done in a quasi-static manner, so that the system is in equilibrium with the thermal bath during the entire process.
For the working medium as a particle of mass $m$  confined in a one dimensional box of length $2a$, the $n^{th}$ energy level is given as
\begin{eqnarray}
E_n = \frac{n^2 \hbar^2 \pi^2}{2m (2a)^2}\quad \text{with}\, n=1,2,3,\dots.
\end{eqnarray}
Therefore, the expression for the partition function $Z_1$ can be written as
\begin{equation}
Z_{(1)} = \sum_{n=1}^{\infty} e^{-\frac{E_n}{k_B T_h}} = \sum_{n=1}^{\infty} e^{-\frac{n^2 \pi^2 \hbar^2}{2m (2a)^2 k_B T_h}}.
\end{equation}
The  partition function $Z_1$ can also be expressed in terms of Jacobi Theta functions as
\begin{equation}
Z_{(1)}= \frac{1}{2} \bigg[-1 + \Theta_3\bigg(0, e^{\frac{\pi^2 \hbar^2}{2m(2a)^2 k_B T_h}}\bigg)\bigg],
\end{equation}
where $\Theta_3(z,q)$ is defined as
\begin{equation} 
\Theta_3(z,q)= 1+ 2\sum_{n=1}^{\infty} q^{n^2} cos(2n z).
\end{equation}

The final state  obtained after the  isothermal process by the quasi-static insertion of the barrier introduces an additional constraint that the probability amplitude of the wave function should be exactly zero in the middle of the box, as well as at the boundaries of the box.
Consequently, the wave functions with odd quantum number are `raised' in energy to the next even numbered energy state. Hence, the energy states with even quantum number are now doubly degenerate. The eigenstates of the new system are given as
\begin{equation}
E_n = \frac{(2n)^2 \pi^2 \hbar^2}{2m(2a)^2}\quad \text{with}\, n=1,2,3,\dots.
\end{equation}
Thus, the partition function can be expressed as
\begin{equation}\label{symins2}
Z_{(2)} = \sum_{n=1}^{\infty} 2 e^{-\frac{(2n)^2\pi^2\hbar^2}{2m(2a)^2k_BT_h}} = 2 Z^a_{T_h},
\end{equation}
where, 
\begin{equation}
\nonumber
Z^a_{T_h}= \sum_{n=1}^{\infty} e^{-\frac{n^2\pi^2\hbar^2}{2m a^2k_BT_h}},
\end{equation}
is the partition function of a particle in a box of length $a$ attached to a bath at temperature $T_h$. In terms of the Jacobi Theta function, the partition function $Z_{(2)}$ of the final state is obtained as
\begin{equation}
Z_{(2)} = 2 \bigg\{ \frac{1}{2} \bigg[-1 + \Theta_3\bigg(0, e^{\frac{\pi^2 \hbar^2}{2ma^2 k_B T_h}}\bigg)\bigg]\bigg\}.
\end{equation}

The heat absorbed in the process is
\begin{equation}
Q_{12}= U_{(2)}-U_{(1)}+ k_{B}T_{h}\ln Z_{(2)}- k_{B}T_{h}\ln Z_{(1)}.
\end{equation}
The change in internal energy is
\begin{equation}
\Delta U_{12} = U_{(2)}-U_{(1)}= -\frac{\partial}{\partial \beta_h} [\ln Z_{(2)} - \ln Z_{(1)}].
\end{equation}
Thus, the work done in the process can be directly written as following  by the first law of thermodynamics:
\begin{equation}
W_{12}= Q_{12} - \Delta U_{12}=k_{B}T_{h}\ln Z_{(2)}- k_{B}T_{h}\ln Z_{(1)}.
\end{equation}

\par
\noindent{\bf Second step\,:}  In this step, the system  is attached with a cold bath at temperature $T_c$ after disconnecting it from the hot bath at temperature $T_h$. This results in lowering the temperature of the system from $T_h$ to $T_c$.

The partition function of the final state can be written as 
\begin{equation}
Z_{(3)} = \sum_{n=1}^{\infty} 2 e^{-\frac{n^2\pi^2\hbar^2}{2m a^2 k_B T_c}}.
\end{equation}
In terms of the Jacobi Theta function, the expression turns out to be:
\begin{equation}
Z_{(3)} = \bigg[ -1 + \Theta_3\bigg(0, e^{\frac{\pi^2 \hbar^2}{2ma^2 k_B T_c}} \bigg)  \bigg].
\end{equation}

 Since no mechanical work is done ($W_{23} = 0$), the heat lost is equal to the change in internal energy:
\begin{equation}
Q_{23} = U_{(3)} - U_{(2)} = -\frac{\partial \ln Z_{(3)}}{\partial \beta_c} + \frac{\partial \ln Z_{(2)}}{\partial \beta_h}.
\end{equation}

\par
\noindent{\bf Third step\,:} In the next step, the barrier is  lifted in a  quasi-static manner such that the system is in  equilibrium with the thermal bath at temperature $T_c$ during the whole process. 
This brings the system to the same energy level structure as that of the initial state $(1)$. The partition function of the final state is given as
\begin{eqnarray}\label{compact}
Z_{(4)} &=& \nonumber \sum_{n=1}^{\infty} e^{-\frac{n^2 \pi^2 \hbar^2}{2m (2a)^2 k_B T_c}} , \\ &=& \frac{1}{2}\bigg[ -1 + \Theta_3\bigg(0, e^{\frac{\pi^2 \hbar^2}{2m (2a)^2 k_B T_c}} \bigg)  \bigg].
\end{eqnarray}
The heat lost at end of the process is obtained in terms of the partition functions as
\begin{equation}
Q_{34}= U_{(4)}-U_{(3)}+ k_{B}T_{c} \ln Z_{(4)}- k_{B}T_{c} \ln Z_{(3)}.
\end{equation}
The work done in the process is
\begin{equation}
W_{34} = Q_{34} - \Delta U_{34} = k_{B}T_{c} \ln Z_{(4)}- k_{B}T_{c} \ln Z_{(3)},
\end{equation}
where $\Delta U_{34}$ is $U_{(4)} - U_{(3)}$.
\newline
\par
\noindent{\bf Fourth step\,:}
In the final step, the system is connected to a thermal bath at temperature $T_h>T_c$ after disconnecting it from the thermal bath at temperature $T_c$. The temperature of the system is raised from $T_c$ to $T_h$ during this process and the system is restored to the initial state $(1)$.
The heat absorbed by the system is given as
\begin{equation}
Q_{41}= -\frac{\partial \ln Z_{(1)}}{\partial \beta_h} + \frac{\partial \ln Z_{(4)}}{\partial \beta_c}.
\end{equation}
These four processes comprise one complete cycle of the heat engine. The net work done by the heat engine in one complete cycle is given as
\begin{eqnarray}\label{pib:workeqn}
W_{\textrm{net}} &=& W_{12} + W_{34},\nonumber  \\ &=& k_{B}T_{c} \ln \frac{Z_{(4)}}{Z_{(3)}} +  k_{B}T_{h} \ln \frac{Z_{(2)}}{Z_{(1)}}.
\end{eqnarray}
Further, the total heat absorbed by the system can be written as
\begin{equation}
Q_{\textrm{in}} = Q_{12} + Q_{41}.
\end{equation}
The efficiency $\eta$ for the cycle is given as:
\begin{equation}\label{pib:effeqn}
\eta = \frac{W_{\textrm{net}}}{Q_{\textrm{in}}}  = 1+\frac{Q_{23}+Q_{34}}{Q_{12}+Q_{41}} .
\end{equation}
More details about PIB-QSHE are available in~\cite{george-pra-2019}.

\subsection{Symmetric insertion of multiple barriers}\label{appendixC}

We now generalize the PIB-QSHE to the scenario where we insert $N$ barriers symmetrically into the box.
To understand the changes in the energy levels, let us consider the insertion of two barriers symmetrically into the box of length $2a$. This divides the box into three equal parts, each of length $2a/3$. The energy levels which have nodes falling on the barrier positions remain unchanged, while the other energy levels shift up to the nearest unchanged energy level. This renders an energy level structure, where each energy state is three-fold degenerate. Similarly, for symmetric insertion of $N$ barriers, we obtain an energy level structure with $N$-fold degeneracy.

 The expression for partition functions at various four stages of the Stirling cycle, as mentioned in Appendix~\ref{appendixA}, for symmetric insertion of $N$ barriers is given as follows:
\begin{eqnarray}\label{multiplep1}
Z_{(1)} &=&  \sum_{n=1}^{\infty} e^{-\frac{n^2 \pi^2 \hbar^2}{2m (2a)^2 k_B T_h}},\nonumber\\&=& \frac{1}{2} \bigg[-1 + \Theta_3\bigg(0, e^{\frac{\pi^2 \hbar^2}{2m(2a)^2 k_B T_h}}\bigg)\bigg].
\end{eqnarray}

\begin{eqnarray}
Z_{(2)} &=&  N\sum_{n=1}^{\infty} e^{-\frac{n^2 \pi^2 \hbar^2}{2m (2a/N)^2 k_B T_h}},\nonumber\\
&=& N\bigg\{\frac{1}{2} \bigg[-1 + \Theta_3\bigg(0, e^{\frac{\pi^2 \hbar^2}{2m(2a/N)^2 k_B T_h}}\bigg)\bigg]\bigg\}.
\end{eqnarray}

\begin{eqnarray}
Z_{(3)} &=&  N\sum_{n=1}^{\infty} e^{-\frac{n^2 \pi^2 \hbar^2}{2m (2a/N)^2 k_B T_c}},\nonumber\\ &=& N\bigg\{\frac{1}{2} \bigg[-1 + \Theta_3\bigg(0, e^{\frac{\pi^2 \hbar^2}{2m(2a/N)^2 k_B T_c}}\bigg)\bigg]\bigg\}.
\end{eqnarray}

\begin{eqnarray}\label{multiplep4}
Z_{(4)} &=&  \sum_{n=1}^{\infty} e^{-\frac{n^2 \pi^2 \hbar^2}{2m (2a)^2 k_B T_c}},\nonumber\\
&=& \frac{1}{2} \bigg[-1 + \Theta_3\bigg(0, e^{\frac{\pi^2 \hbar^2}{2m(2a)^2 k_B T_c}}\bigg)\bigg].
\end{eqnarray}
 Using these partition functions, we can write the corresponding work~(\ref{pib:workeqn}) and efficiency equation~(\ref{pib:effeqn}).

\subsection{Asymmetric insertion of single barrier}\label{appendixB}

We insert a barrier asymmetrically into the box such that it is divided into two parts of length $x$ and $y$ with $x+y=2a$.

The energy eigenstates for a particle in a box of length $x$ and $y$ is given as follows:
\begin{eqnarray}
    E_i = \sum_{i=1}^{\infty} e^{-\frac{i^2 \pi^2 \hbar^2}{2m x^2 }} \quad \text{with} \,i=1,2,3,\dots,
\end{eqnarray}
and
\begin{eqnarray}
    E_i' = \sum_{i=1}^{\infty} e^{-\frac{{i'}^2 \pi^2 \hbar^2}{2m y^2 }} \quad \text{with}\, i'=1,2,3,\dots.
\end{eqnarray}
 The energy level structure is a collection of all these energy eigenstates corresponding to the two boxes. 
Therefore, the partition function after the  asymmetric  insertion of barrier is the sum of Boltzmann factor over the whole energy level structure, \ie, all the energy eigenstates corresponding to the two boxes.

The expression for various partition functions at the four stages of the Stirling cycle,  as discussed in Appendix~\ref{appendixA}, for asymmetric insertion of a single barrier is given as follows:
\begin{eqnarray}\label{asymp1}
Z_{(1)} &=&  \sum_{n=1}^{\infty} e^{-\frac{n^2 \pi^2 \hbar^2}{2m (2a)^2 k_B T_h}},\nonumber\\ &=& \frac{1}{2} \bigg[-1 + \Theta_3\bigg(0, e^{\frac{\pi^2 \hbar^2}{2m(2a)^2 k_B T_h}}\bigg)\bigg].
\end{eqnarray}

\begin{eqnarray}\label{asymp2}
Z_{(2)} &=&  \sum_{n=1}^{\infty} e^{-\frac{n^2 \pi^2 \hbar^2}{2m x^2 k_B T_h}}+ \sum_{n=1}^{\infty} e^{-\frac{n^2 \pi^2 \hbar^2}{2m y^2 k_B T_h}},\nonumber\\
\nonumber &=& \frac{1}{2} \bigg[-1 + \Theta_3\bigg(0, e^{\frac{\pi^2 \hbar^2}{2mx^2 k_B T_h}}\bigg)\bigg]\\ & &
+\frac{1}{2} \bigg[-1 + \Theta_3\bigg(0, e^{\frac{\pi^2 \hbar^2}{2my^2 k_B T_h}}\bigg)\bigg].
\end{eqnarray}

\begin{eqnarray}
Z_{(3)} &=&  \sum_{n=1}^{\infty} e^{-\frac{n^2 \pi^2 \hbar^2}{2m x^2 k_B T_c}}+ \sum_{n=1}^{\infty} e^{-\frac{n^2 \pi^2 \hbar^2}{2m y^2 k_B T_c}},\nonumber\\
\nonumber &=& \frac{1}{2} \bigg[-1 + \Theta_3\bigg(0, e^{\frac{\pi^2 \hbar^2}{2mx^2 k_B T_c}}\bigg)\bigg]\\ & &
+\frac{1}{2} \bigg[-1 + \Theta_3\bigg(0, e^{\frac{\pi^2 \hbar^2}{2my^2 k_B T_c}}\bigg)\bigg].
\end{eqnarray}

\begin{eqnarray}\label{asymp4}
Z_{(4)} &=&  \sum_{n=1}^{\infty} e^{-\frac{n^2 \pi^2 \hbar^2}{2m (2a)^2 k_B T_c}},\nonumber\\
&=& \frac{1}{2} \bigg[-1 + \Theta_3\bigg(0, e^{\frac{\pi^2 \hbar^2}{2m(2a)^2 k_B T_c}}\bigg)\bigg].
\end{eqnarray}
 The above partition functions can be utilised to write the work~(\ref{pib:workeqn}) and efficiency equation~(\ref{pib:effeqn}).
 We note  that for $x=y$, the situation becomes the same as the symmetric insertion of  a single barrier case, which has been dealt with in Appendix~\ref{appendixA}.


\begin{thebibliography}{55}
\bibitem [{tay()}]{taylor-harmonic}%
	\BibitemOpen
	\href@noop {} {}\bibinfo {note} {Any generic potential can be Taylor expanded
		near equilibrium at $x=x_0$ as $U(x) =
		U(x_0)+(x-x_0)\frac{dU}{dx}|_{x=x_0}+\frac{(x-x_0)^2}{2!}\frac{d^2U}{dx^2}|_{x=x_0}+\cdots$.
		Since at equilibrium, $\frac{dU}{dx}|_{x=x_0}$ is zero and usually the
		physics is not affected by the constant potential $U(x_0)$. Therefore, we can
		approximate the generic potential as harmonic oscillator, $U(x) \approx
		\frac{(x-x_0)^2}{2!}\frac{d^2U}{dx^2}|_{x=x_0} $, for small oscillations
		around equilibrium.}\BibitemShut {Stop}%
\end{thebibliography}

%

\end{document}